\documentclass[11pt,twoside]{article}

\usepackage{asp2006}
\usepackage{epsf}
\usepackage{psfig}
\usepackage{lscape}

\begin{document}
\title{New Insights into the Quasar Type1/Type2 Dichotomy from Correlations between Quasar Host Orientation and Polarization}   
\author{B. Borguet, D. Hutsem\'{e}kers, G. Letawe, Y. Letawe and P. Magain}   
\affil{IAGL, University of Li\`{e}ge, All\'{e}e du 6 Ao\^{u}t, 17, 4000 Li\`{e}ge, Belgium}    

\begin{abstract} 
 We investigate correlations between the direction of the optical linear polarization and the orientation of the host galaxy/extended emission for type1 and type2 radio-loud and radio-quiet quasars. We have used high resolution Hubble Space Telescope data and a deconvolution process to obtain a good determination of the host galaxy/extended emission (EE) position angle. With these new measurements and a compilation of data from the literature, we find a significant correlation, different for type1 and type2 objects, between the linear polarization position angle and the orientation of the EE, suggesting scattering by an extended UV/blue region in both types of objects. Our observations support the extension of the Unification Model to the higher luminosity AGNs like the quasars, assuming a two component scattering model.
\end{abstract}

 \keywords{quasars: general -- polarization}



\section{Introduction}

 The discovery of type1-like broad emission lines in the polarized spectrum of the type2 Seyfert galaxy NGC1068 \citep{borguet_anto85} led to the foundation of the so-called Unification Model (UM) of AGNs \citep[e.g.][]{borguet_anto93} in which the orientation of a dusty torus plays a crucial role in the spectroscopic classification of the AGN for a given observer. While this UM is now well accepted for the lower luminosity AGNs, a key question is whether it applies to higher luminosity objects like quasars since the presence of the dusty torus may be affected by the higher radiation flux.
 
 In order to get some insights into the quasar inner structure, we investigate the possible existence of a correlation between the direction of the linear optical polarization ($\theta_{pola}$) and the orientation of the major axis ($PA_{host}$) of the host galaxy/extended emission surrounding RQ and RL quasars. 
 
 \section{Analysis and Interpretation}
 
 The sample used in our study is made of type1 and type2 RQ/RL quasars for which high resolution visible/near-IR images are available and with optical polarization data reported in the literature. We determined the $PA_{host}$ for the objects with missing measurements by using the MCS deconvolution method \citep*{borguet_mag98}. From this compilation, we selected a sub-sample of objects for which we have relevant and accurate data and then computed the acute angle $\Delta \theta$ between the directions defined by the $PA_{host}$ and $\theta_{pola}$ angles. Further details about the quasar sample and the determination of the $PA_{host}$ are available in \citet{borguet_bo08}.
 
 \begin{figure}
 \plottwo{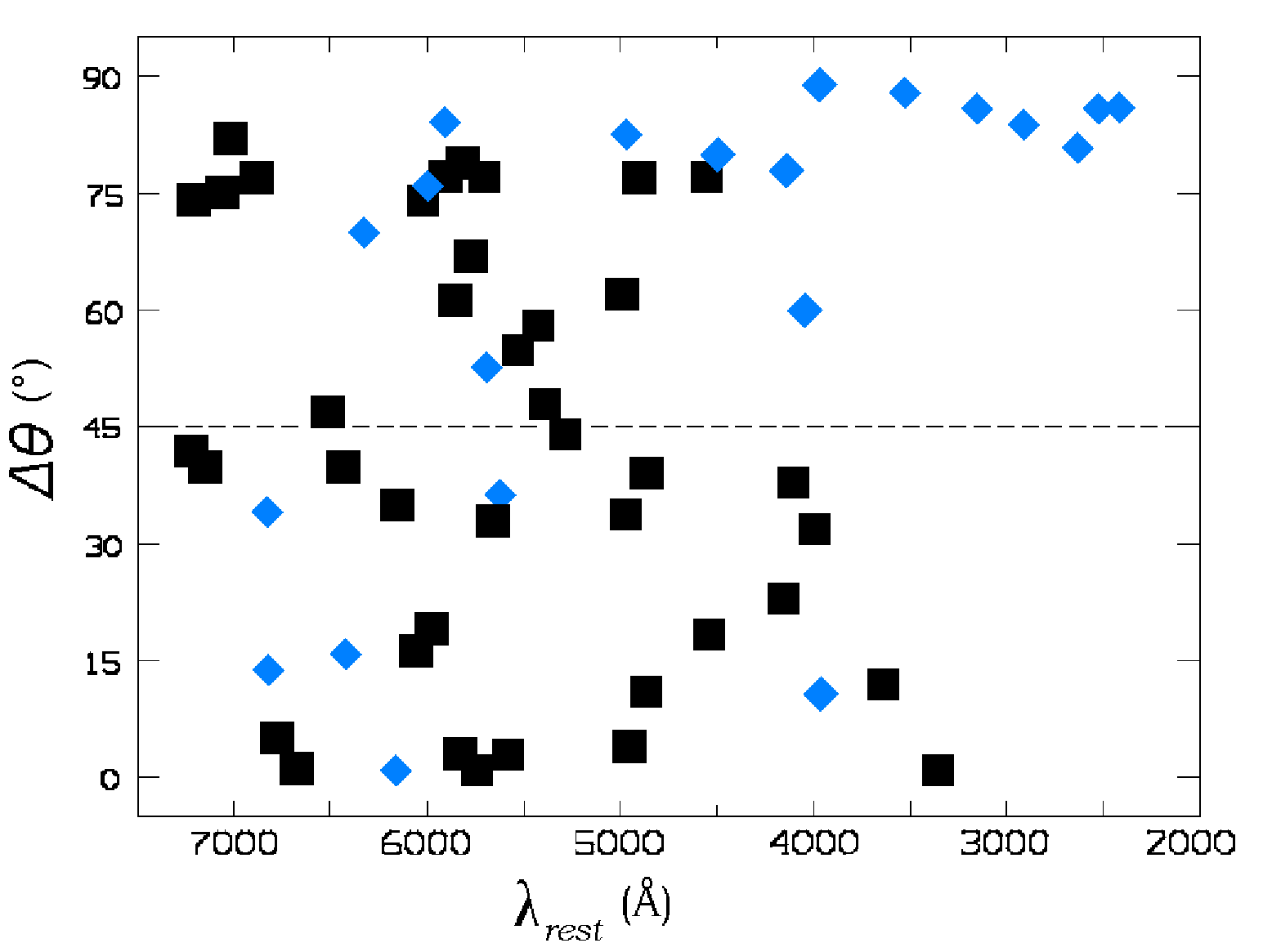}{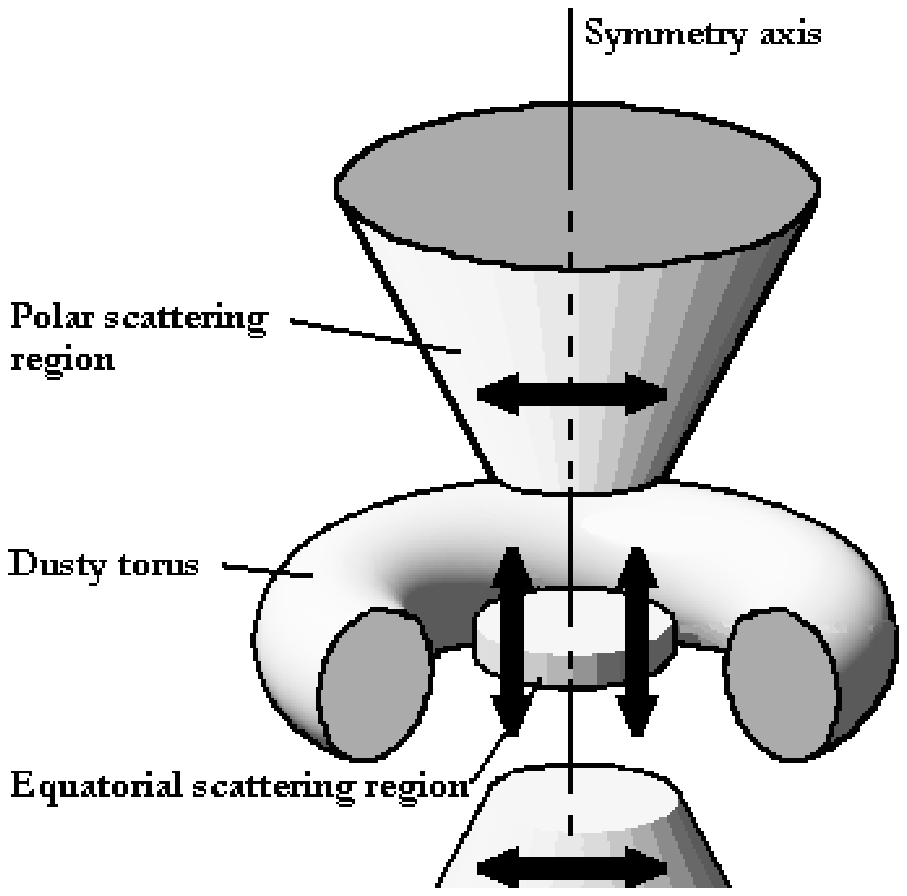}
 \caption{Left : distribution of the acute angle $\Delta \theta$ as a function of $\lambda_{rest}$ (see text). Type1 objects are pictured by squares and type2 objects by diamonds. Right : illustration of the two component scattering model (see text, adapted from \citet{borguet_smi04}). The black arrows illustrate the polarization direction produced in each scattering region. Type1 and type2 objects are seen under different viewing angle.}
 \end{figure}
 
 The behavior of the angle $\Delta \theta$ as a function of the observation wavelength $\lambda_{rest}$ as measured in the quasar rest-frame is illustrated in the left panel of Fig.~1. From the analysis of the data plotted, we conclude that while no particular behavior is noted at redder wavelengths ($\lambda_{rest}$ $\geq 5000$ \AA), a clear dichotomy appears in the UV/blue domain where the type1 quasars tend to have their EE preferentially aligned ($\Delta \theta \leq 45\deg$) with the polarization direction, these two axes being mainly orthogonal ($\Delta \theta \geq 45\deg$) in the type2 objects. Moreover, the observed alignment effect seems to be independent of the radio-loudness.
 
 In the case of the type2 quasars, this behavior is known as the \textit{alignment effect} reported by \citet{borguet_ci93} and \citet{borguet_za06} where the extended UV/blue emission resolved in images is interpreted as an electron/dust polar region, scattering off the nuclear light and explaining the anti-alignment. In type1 quasars, the alignment can be explained by assuming a two-component scattering model (see Fig.~1), similar to the one proposed for the Seyfert galaxies by \citet{borguet_smi04}. In this scenario, an equatorial scattering region located inside the dusty torus produces a polarization aligned with the torus symmetry axis. The resulting polarization, the sum of the polar and equatorial contributions, is dominated by the equatorial component because of the higher symmetry of the polar region at smaller viewing angles \citep[see][]{borguet_bo08}.



\end{document}